\documentclass[runningheads]{llncs}
\usepackage{graphicx}
\usepackage[T1]{fontenc} 
\usepackage[utf8]{inputenc}
\usepackage{amsmath}
\usepackage{caption} 
\usepackage{wrapfig}
\usepackage{outlines} 
\usepackage[ruled,vlined]{algorithm2e}
\usepackage{tabularx}
\usepackage[table]{xcolor}
\usepackage{xltabular}
\usepackage{booktabs}
\usepackage{multirow}
\usepackage[misc,geometry]{ifsym}
\graphicspath{ {images/} }
\begin{document}
\title{A Probabilistic Approach to Personalize Type-based Facet Ranking for POI Suggestion}
\titlerunning{A Prob. Approach to Personalize T-Facet Ranking}

%
\author{Esraa Ali \inst{1} \Letter \orcidID{0000-0003-1600-3161} \and
Annalina Caputo \inst{2} \Letter \orcidID{0000-0002-7144-8545} \and
Séamus Lawless\inst{1}\orcidID{0000-0001-6302-258X} \and
Owen Conlan\inst{1}\orcidID{0000-0002-9054-9747}
}
\authorrunning{E. Ali et al.}
%
\institute{ ADAPT Centre, School of Computer Science and Statistics, Trinity College Dublin
\email{esraa.ali,seamus.lawless,owen.conlan@adaptcentre.ie}\\
 \and
ADAPT Centre, School of Computing,  Dublin City University\\
\email{annalina.caputo@adaptcentre.ie}}
\maketitle              
\begin{abstract}

Faceted Search Systems (FSS) have become one of the main search interfaces used in vertical search systems, offering users meaningful facets to refine their search query and narrow down the results quickly to find the intended search target. This work focuses on the problem of ranking type-based facets.  In a structured information space, type-based facets (t-facets) indicate the category to which each object belongs. When they belong to a large multi-level taxonomy, it is desirable to rank them separately before ranking other facet groups. This helps the searcher in filtering the results according to their type first. This also makes it easier to rank the rest of the facets once the type of the intended search target is selected. Existing research employs the same ranking methods for different facet groups. In this research, we propose a two-step approach to personalize t-facet ranking. The first step assigns a relevance score to each individual leaf-node t-facet. The score is generated using probabilistic models and it reflects t-facet relevance to the query and the user profile. In the second step, this score is used to re-order and select the sub-tree to present to the user. We investigate the usefulness of the proposed method to a Point Of Interest (POI) suggestion task. Our evaluation aims at capturing the user effort required to fulfil her search needs by using the ranked facets. The proposed approach achieved better results than other existing personalized baselines.

\end{abstract}
\section{Introduction}
In Faceted Search Systems (FSS), users explore the information space through facets, which are attributes or meta-data that describe the underlying content of the collection. As the magnitude of data in a collection increases, the number of facets and their values becomes impractical to display on a single page. Providing users with too many facets has been shown to overwhelm and distract them \cite{koren2008personalized}. 

Faceted browsers overcome this problem by either displaying a small number of facets and making the rest accessible through a ``more'' button, or by displaying only the facet titles without the values: if the user is interested in a facet they can click on the title to view its values. 
In either case, ranking the top facets is required as it assists the searcher in narrowing down the information space and locate the target document with minimum effort.

In an information space that is structured, facets are either extracted from the edges or relationships between objects, in which case they are called \textit{property-based facets (p-facets)}, or they are extracted from the types of the objects, in which case they are called \textit{type-based facets (t-facets)} (e.g. values of subClassOf or isA relationships). 
Systems vary in their use of facets, some use a single type of facets, others mix the two types. 
Usually, this is done by presenting t-facets first, followed by the p-facets \cite{FacetsSurvey:2016}.
In FSS, which exploit multiple types of resources\footnote{In the scope of this paper, we refer to  resources (or information objects) being searched as venues or POIs.}, it is important to prioritize and focus  on the relevant t-facets.
This is especially true when the types of resources come from a large multilevel hierarchy. 
This will encourage the user to filter the results by their type first and  make it easier to  rank the p-facets. Multilevel hierarchical types are derived from ontologies by exploiting the \texttt{subClassOf} relationships.

In this work, we focus on analysing the role of personalization in t-facet ranking in isolation from other FSS aspects. 
Existing facet ranking methods rely on attribute frequencies, navigation cost models, textual queries or click logs to order the facets \cite{FacetsSurvey:2016}. 
Neither the special case of t-facet ranking nor the fact that t-facets relevance can be user dependant are addressed by these approaches. This experiment aims at answering the following research question: \emph{$\textbf{RQ}$:  Does personalizing the t-facet ranking using probabilistic scoring models minimize users effort to fulfil their search needs?}

This study contributes to the research in this area by introducing a novel ranking algorithm for type-based facets. The algorithm exploits the user's past preferences to build a user profile for ranking type-based facets.
The proposed approach functions over two consecutive steps. The first step generates personalized relevance score for each t-facet at the end level of the taxonomy. Then, the second stage aggregates this score to re-arrange the ancestor t-facet nodes and re-build the final t-facet tree to be rendered to the user. To the best of our knowledge, this is the first approach that focuses on the special case of t-facet ranking. It investigates using topic-based user profiles to improve the ranking process. In addition to that, it provides an effective strategy to rank different t-facet levels and decide the final tree to be portrayed to the searcher. The approach operates on t-facets, which have a well structured tree-like hierarchical taxonomy.

The implemented approach is evaluated using the TREC Contextual Suggestion (TREC-CS) track dataset \cite{TRECCS:2016}. 
TREC-CS is a personalized Point-Of-Interest (POI) recommendation task, in which participants develop systems to give a ranked list of suggestions related to a given user profile and a context. We solve the POI suggestion problem by ranking the types of venues as t-facets. In our evaluation, we measure the extent to which this ranked tree minimizes the user effort to reach the first relevant POI.

\section{Facet Ranking Related Research}
\label{sec:sota}
Several approaches have been proposed in the literature to solve the problem of personalized facet ranking that make use of individual user models, collaborative filtering (CF), or a mixture between the two.
Factic is a FSS that personalizes by building models from semantic usage logs.
Several layers of user adaption are implemented and integrated with different weights to enhance the facet relevance model \cite{tvarovzek2010factic}. 
Koren et al. \cite{koren2008personalized} suggested a CF approach by leveraging explicit user feedback about the facets, which is used to build a facet relevance model for individuals.
They also use the aggregated facet ratings to build a collaborative model for the new users in order to provide initial good facets in absence of a user profile. 
The Adaptive Twitter search system generates user models from Twitter to personalize facet-values ordering  \cite{AdaptiveTwitter:2011}.
The user model contains entities extracted from the user's tweets. 
The facet-values are weighted higher if they exist in the user profile. 
Le et al. \cite{Le2012} also collects user profile from social networks. 
The profile is learned from user activities and preferences using a tf-idf feature vector model. Important facets are then highlighted through a matching with the model. 
A personalized ranking based on CF features was suggested by Chantamunee et. al \cite{Chantamunee2018}.

They used user ratings and Matrix Factorization via SVM to learn facet ranks. 
All the discussed approaches in this section use the same strategy to rank p-facets and t-facets. They do not provide means to order the hierarchy of t-facets.
We believe it is important to distinguish between the two types of facets during the ranking process as they each support the user in different ways in finding their intended target. Our approach exploits topic-based user profiles, which employs users' historical ratings to infer their preferred t-facet, an area which was not explored by earlier research in facet ranking.

\section{Proposed Approach}

Our method works in the context of personalized venue search. When a user submits a query, the underlying search engine retrieves a relevant set of venues for it\footnote{How the venue ranking is performed is outside scope of this research.}. Our method works on this set by collecting the t-facets associated with the retrieved venues. We assume that this set of retrieved venues is relevant for the query and can be considered as the input for the t-facet ranking algorithm.

The proposed t-facet ranking approach consists of two steps.  Assuming that the t-facets are organized in a taxonomy, the first step assigns a relevance score to each t-facet leaf node. The input to this step is the retrieved venues with their relevancy score, the t-facets to which they belong, as well as the user profile. 
The second step constructs the final t-facet tree to be displayed to the user. The input to this step is both the score for each t-facet (generated at the first step) and the original hierarchical taxonomy from which we derived the t-facets. The output of the t-facet ranking is a sub-tree which contains the ordered set of relevant t-facets. 
The following sub-sections provide the details of each step.

\subsection{Step 1: Scoring Using T-Facet Probabilistic Models }
\label{sec:prob-ptfs} 

In this step, probabilistic models are developed to estimate a t-facet relevance score given a query and a user profile.  The models are based on the well-known probabilistic models introduced by Sontag et al. for personalized web search \cite{Probabilistic:2012}. They personalized the search results using  topic-based user profiles collected from users' historical interactions with  the system. Their approach re-weights the original document level search results according to topic relevancy to the user and query. In this work, we utilize the topic re-weighting factor to derive the t-facets score. The generated score reflects t-facet relevance to both the user and the input query. Below we re-define those models in the context of our t-facet scoring task. 
To generate the t-facet score, two models are proposed; \emph{\textbf{Model-1}} assumes no background data available,
it is calculated using the following formula:
\begin{equation}
    score(f_i) = \sum_{f_u} P(f_u|q,\theta_u) \times P(cov(f_u,f_i)|f_u,f_i)
\end{equation}
Where $f_i$ is the current t-facet to be ranked, $f_u$ are t-facets rated before by the user, $\theta_u$ is the user profile, and  $P(cov(f_u,f_i)|f_u,f_i)$ is the probability that t-facet $f_u$ is \emph{covered} by the t-facet $f_i$. We estimate this probability using two methods, the first is the \emph{\textbf{exact}} match, in which the probability equals to  1 if $f_u=f_i$, 0 otherwise. The second estimate (\emph{\textbf{cosine}}) uses a function of distance between $f_u$ and $f_i$. In our case, we employ the cosine similarity between BERT vectors generated for the input t-facet labels, using a pre-trained generic BERT model.
Details about the probability $P(f_u|q,\theta_u)$ are provided later in this section.  

\emph{\textbf{Model-2}} uses background data to estimate the score: 

\begin{equation}
    score(f_i)= \frac{\sum_{f_u}  P(f_u|q,\theta_u) \times P(cov(f_u,f_i)|f_u,f_i)}{\sum_{f}P_r(f|q) \times P(cov(f,f_i)|f,f_i)}
\end{equation}
The numerator is the same as Model-1. In the denominator, the background distribution $P_r(f|q)$ (where $r$ denotes a random or generic user) is calculated by averaging the relevance score for the top $N$ search results belonging to this t-facet when the query $q$ is submitted to search engine.
It can be obtained using the following equation:
\begin{equation}
\label{eq:p_f_q}
     P_r(f|q) = \frac{1}{N} \times \sum_{m=1}^{N} P(rel(d_m,q)=1|q) \times P(f_d|d_m)
\end{equation}
Note that in our case $P(f_d|d)=1$, since the venues' types are assigned by their owners, i.e. the type of the venue is not estimated it is given, and hence can be dropped from (\ref{eq:p_f_q}). Details of the derivation of the models can be found in \cite{Probabilistic:2012}.

To model the user's preferences, we estimate $P(f_u|q,\theta_u)$, for which we use users historical ratings by assuming that users prefer t-facets of the venues they rated positively in the past. We use the generative model suggested in \cite{Probabilistic:2012} to estimate this value by employing the Bayesian rule:
\begin{equation}
   P(f_u|q,\theta_u)=\frac{P(f_u|\theta_u) \times P(q|f_u)}{\sum_{f'}P(f'|\theta_u) \times P(q|f')}
\end{equation}

$P(f_u|\theta_u)$ is estimated by dividing how many times the user rated documents belong to $f_u$ positively, divided by total number of documents rated by the user.
The probability  $P(q|f)$  is estimated by inverting  $P(f|q)$ (see Eq.\ref{eq:p_f_q}):
\begin{equation}
     P(q|f) = c \frac{P_r(f|q)}{P_r(f)}
\end{equation}
Where $c$ is a constant and $P_r(f)$ is the probability that a random user rates this facet positively obtained by counting how many times this t-facet's documents were rated positively by all users divided by the total number of rated documents.

\subsection{Step 2: T-Facet Tree Building}
\label{sec:construction}

The tree construction algorithm re-orders the original taxonomy tree by using the generated scores from the previous step. It follows a bottom-up approach where the t-facets at the lower level in the taxonomy are sorted first, then it proceeds by sorting all the ancestors of those t-facets, and so on up to the root of the hierarchy.  At each level, the scores from the previous level are employed to induce the ranks of the current level. 
This step also decides which top t-facets sub-tree will appear to the user in the first result page. Remaining facets will be available to the user by clicking 'More' link.\footnote{Although we acknowledge that other HCI factors may influence the decision of what portion of the tree should be displayed to the users, in this work we focus on studying how the tree building approach affects the user from a pure metric perspective.}

\begin{figure}[h]
\includegraphics[width=\textwidth]{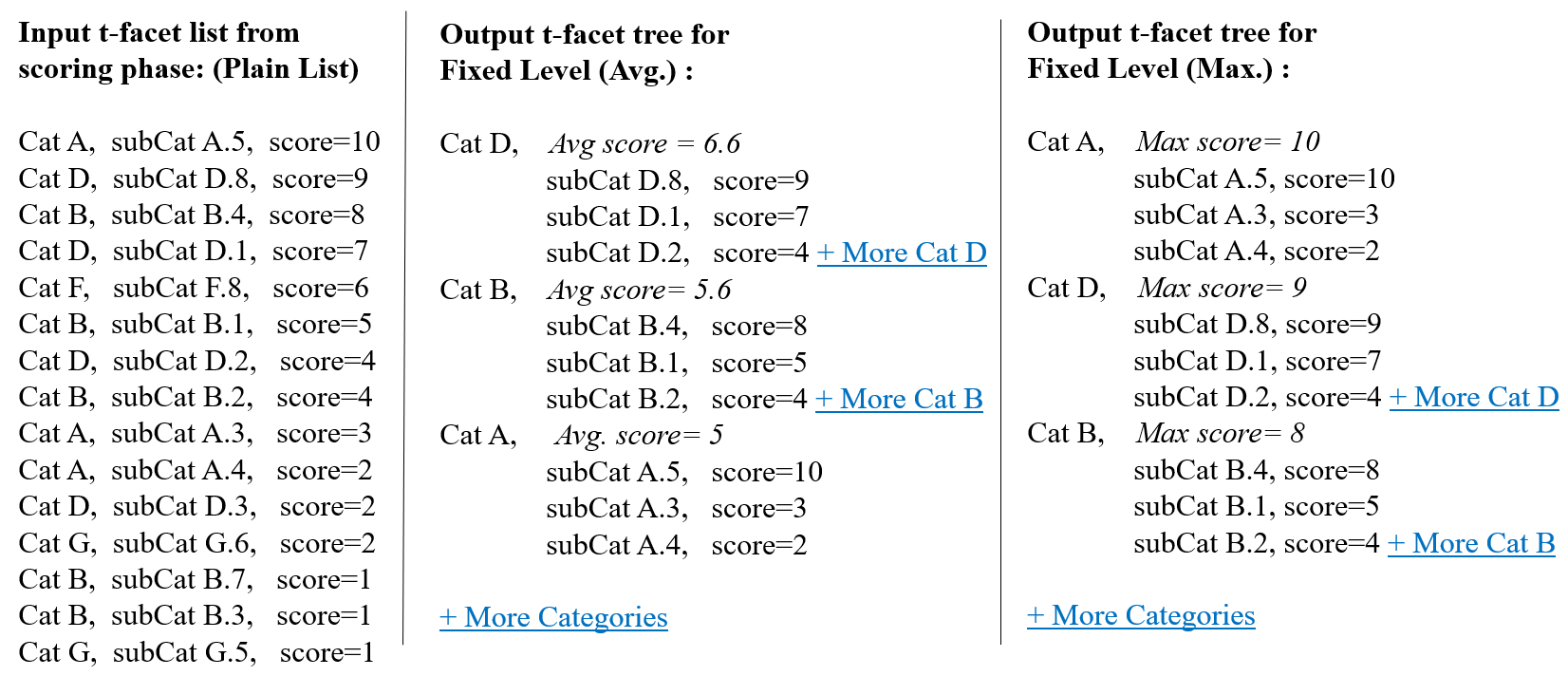}
\caption{Example output t-facet tree for a 2 level taxonomy using two aggregations: average and max, level-1 t-facet page size=3, level-2 t-facet page size=3.}
\label{fig:tree-plain-grouping}
\end{figure}

To build a final t-facet tree with $v$ levels, we adopted a \emph{\textbf{fixed level}} strategy that follows a bottom-up approach. The strategy respects the original taxonomy hierarchy and uses a predefined fixed page size for each t-facet level. The strategy starts by grouping t-facets at level-$v$ by their parent. Then, it sorts the (parent) nodes at level-$(v-1)$ by aggregating the scores of their top $k$ children, the children are ordered by their relevance score generated in step 1, and so on up to level-1. Several aggregation functions can be used, in our experiments we used average (Avg) and maximum (Max) functions.
Figure \ref{fig:tree-plain-grouping} shows an example for this process, categories (Cat.) correspond to level-1 t-facets and sub-categories correspond to level-2 t-facets. In the case where a level-1 facet has additional relevant t-facet children that are not displayed in first page, they will be available to the user through the "+ More Cat ...". Each following t-facet page will be sorted in the same way.  The final output provides the user a more organized and readable t-facet tree.

\section{Experimental Results}

\paragraph{Experimental setup.} Our approach is evaluated on TREC-CS 2016 dataset \cite{TRECCS:2016}.  The t-facet taxonomy is derived from the Foursquare venue category hierarchy~\footnote{https://developer.foursquare.com/docs/resources/categories, version: 20180323}. Hence, having as much Foursquare venues linked to TREC-CS POIs as possible is paramount. 
For this reason, we complement the original data with three Foursquare supplementary datasets from  \cite{Aliannejadi:2017,bayomi2016adapt_tcd} and our own crawled  POIs. The final dataset has 58 requests and an average of 208 t-facets per request to be ranked. We consider the first two levels of the taxonomy, they contain 10 level-1 and  429 level-2 t-facets. The document search engine implements BM25 with NDCG value of 0.4023, the query is formed by combining user weighed tags by their most common rating.

The existence of relevance judgments makes it possible to evaluate our approach against a well established ground-truth. We follow the evaluation approach used in Faceted Search task of INEX 2011 Data-Centric Track\cite{inex2011}. 

We report two metrics suggested by task organizers. The number of actions (\#Actions) metric counts how many clicks the user has to perform on the ranked facets list in order to reach the first relevant document in the top 5 results. The faceted scan (F-Scan) metric measures the user's effort to scan facets and documents until they reach the same document. We focus on these two metrics as a proxy for user's effort, which will help in answering our research question.

We report the results for the no background model (Model-1) and the background model (Model-2), each experimented using two coverage probability estimators (exact) and (cosine). To show the effect of different tree building approaches on the evaluation metrics, we produce results using two strategies: 1) fixed level with average (Avg), 2) fixed level with maximum (Max). Both use 3 level-1 t-facets per page, with 3 level-2 t-facets each.
\paragraph{Preliminary Results.}

The overall results in Table~\ref{tab:fixed_max} show that the no Model-1 consistently outperforms  Model-2 across all metrics and regardless the used coverage probability method. One possible explanation is the small number of training points available for the estimation of the background model. The dataset has only 26 users, each rated either 30 or 60 venues, and the same 60 venues are rated by all users. 
As a result, the profiles are limited to a small set of t-facets, which ultimately affected the probability distributions.  
Further in depth analysis of the relation between the number of user historical POIs and the performance of the scoring methods is needed. 
\begin{table}
\vspace{-.5cm}
\centering
\caption{\textbf{Results for Probabilistic Scoring and Fixed Level Tree Building Strategy with Max and Avg Aggregation Functions.}}\label{tab:fixed_max}
\setlength{\tabcolsep}{1.2em}
\begin{tabular}{ l r l r l  }
\toprule
\multirow{2}{*}{Scoring Method} & \multicolumn{2}{c}{\text{Max}} & \multicolumn{2}{c}{\text{Avg}}\\
  & F-Scan & \#Actions & F-Scan & \#Actions\\
\midrule
 Model-1 + exact & 4.258&1.534 & 4.051 & 1.517 \\
Model-1 + cosine & \textbf{3.413} &\textbf{1.327} & \textbf{3.482} & \textbf{1.396}  \\
 Model-2 + exact & 4.534 &1.706 & 4.327 & 1.706 \\
 Model-2 + cosine & 4.844 &1.758 & 4.879 & 1.810 \\
\bottomrule
\end{tabular}
\vspace{-.3cm}
\end{table}

The skewed t-facet probability distribution also explains why the \textit{cosine} similarity implementation gave better results in Model-1. It aids the score generation for new unseen, t-facets, where the strict \textit{exact} match approach fails to handle such cases, since it assigns 0 score if the user never rated that category before. 

From table \ref{tab:fixed_max} we can also observe that the evaluation metrics were affected by the used tree building strategy. For the best performing scorer (Model-1 + cosine), the Fixed Level-Max strategy produced better results. Two factors played a role here: 1) The strategy maintains the top scored level-2 facets at the top of the final tree; 2) In estimating $P_r(f|q)$ (see Eq.\ref{eq:p_f_q}) we set $N=1$ (for all models) to favor t-facets which will promote the first relevant result early to the user, which in turn effectively minimized the user effort as shown in the results. When experimenting with higher $N$ values, all metrics were negatively impacted. 
\begin{wraptable}{r}{7cm}
\vspace{-.5cm}
\centering
\caption{\textbf{Comparing our results against baselines using Fixed Level-Max}}\label{tab:baselines}
\setlength{\tabcolsep}{0.3em}
\begin{tabular}{ lrr }
\toprule
 Scoring Method & F-Scan & \#Actions \\
\midrule
 Model-1 + cosine & 3.413 &\textbf{1.327}\\
 MF-SVM \cite{Chantamunee2018} &3.741 & 1.431 \\
 Most Prob. (Person) \cite{koren2008personalized}& 4.000 &1.672 \\
 Most Prob. (Collab) \cite{koren2008personalized}& \textbf{3.327} & 1.379\\
\bottomrule
\end{tabular}
\vspace{-.5cm}
\end{wraptable}

Table \ref{tab:baselines} compares our system performance against three personalized facet ranking approaches.
Since none of the existing methods handle the hierarchical nature of the t-facets, we use them as scoring methods with the Fixed Level-Max strategy. We can see that our Model-1+cosine scoring method achieved minimum \#Actions.  The Most Prob. (Collab) approach achieved competing results, with F-Scan slightly better than our model. A reason for this result is that by favouring popular t-facets, this method worked well given the skewed t-facet probability distribution. However, it has the disadvantage of not handling new unseen t-facets and failing for users with unpopular preferences. Our approach on the other hand, handles both cases effectively.

\section{Conclusions}
This work has introduced a novel t-facet ranking approach. The two-step approach  considers the hierarchical nature of t-facets as well as user individual preferences. The first step assigns score to t-facets. The second step uses the score to re-arrange and build the final t-facet tree to the user. To personalize the scores, we explored several probabilistic models. They have shown promising results given the limited user profiles in the dataset. Our future plans include experimenting with more POI suggestion datasets and experimenting with  complex taxonomies to better understand the behavior of the proposed methods.
Our experiments have demonstrated that even the  straight-forward tree building approaches can aid the ranking process. Developing more advanced strategies can introduce further improvement.

\subsubsection*{Acknowledgements}
This research was conducted with the financial support of Science Foundation Ireland (SFI) under Grant Agreement No. 13/RC/2106 at the ADAPT SFI Research Centre at Trinity College Dublin.  The ADAPT  Centre for Digital Media Technology is funded by SFI through the SFI Research Centres Programme and is co-funded under the European Regional Development Fund (ERDF) Grant No. 13/RC/2106\_P2.
\bibliographystyle{splncs04}
\bibliography{tfacet-bib}

\end{document}